\documentclass[10pt,showpacs,showkeys,twocolumn]{revtex4-1}
\usepackage{bm}
\usepackage{amssymb}
\usepackage{amsmath}
\usepackage{graphicx}
\usepackage{multirow}
\usepackage{MnSymbol}
\usepackage{wasysym} 
\usepackage{stmaryrd}
\usepackage[outline]{contour}
\usepackage[dvipsnames]{xcolor}
\usepackage{float}

\definecolor{olive}{RGB}{182,187,37}
\usepackage[mathlines]{lineno}

\begin{document}

\title[Scaling rules for the ionization of biological
molecules]{Scaling rules for the ionization of biological
molecules by highly charged ions}
\author{A. M. P. Mendez, C. C. Montanari, J. E. Miraglia}
\affiliation{Instituto de Astronom\'{\i}a y F\'{\i}sica del Espacio, 
Consejo Nacional de Investigaciones Cient\'ificas y T\'ecnicas - 
Universidad de Buenos Aires, Pabll\'on IAFE, 1428 Buenos Aires, 
Argentina}

\begin{abstract}
We investigate scaling rules for the ionization cross sections of
multicharged ions on molecules of biological interest. The cross 
sections are obtained using a methodology presented in [Mendez 
\textit{et al.} J. Phys B (2020)], which considers distorted-wave 
calculations for atomic targets combined with a molecular 
stoichiometric model. We examine ions with nuclear charges $Z$ from 
$+1$ to $+8$ impacting on five nucleobases --adenine, cytosine, 
guanine, thymine, uracil--, tetrahydrofuran, pyrimidine, and water. 
We investigate scaling rules of the ionization cross section with 
the ion charge and the number of active electrons per molecule. 
Combining these two features, we define a scaling law for any ion and 
molecular target, which is valid in the intermediate to high energy 
range, i.e., 0.2-5 MeV/amu for oxygen impact. Thus, the forty 
ion-molecule systems analyzed here can be merged into a single band. 
We confirm the generality of our independent scaling law with several
collisional systems. 
\end{abstract}


\maketitle

\section{Introduction}

The interest in the ionization of biological molecules by 
multicharged ions has increased due to medical and environmental
implementations~\cite{PhysMed}, including medical 
treatments~\cite{Mohamad2017,Solov2009,Denifl2011} and contaminant 
recognition in biological materials~\cite{water,ferrazdias}. 
Many semiempirical \citep{vera_prl2013} and theoretical efforts are 
currently being undertaken~\cite{MendezJPB20,Quinto20,ludde2019,
ludde2018,ludde2016,Champion2012} to get reliable values for the 
ionization cross sections of these molecular systems. 

In recent work~\cite{MendezJPB20}, we combined the continuum 
distorted-wave calculations (CDW) for atoms and the simple 
stoichiometric model (SSM) to approximate the ionization cross 
sections of complex molecular targets by the impact of charged ions. 
The molecular ionization cross section~$\sigma_M$ was expressed as 
a linear combination of atomic CDW calculations $\sigma_A$, 
weighted with the number of atoms for each specie $n_A$, i.e, 
$\sigma_M=\sum_A n_A \sigma_A$. The CDW-SSM approximation showed 
consistent results for over a hundred of biologically relevant 
ion-molecule systems. As expected, in the high energy range (i.e., 
above 5 MeV/amu), the ionization cross sections of 
the molecular systems follow the $Z^2$ dependence predicted by the 
first Born approximation. However, at intermediate 
energies, the dependence with $Z$ is not straightforward since 
non-perturbative models are mandatory.

This contribution constitutes a follow-up to our previous 
work~\cite{MendezJPB20}. We introduce here a two-folded scaling rule 
for the ionization cross sections of complex molecules by charged ions. 
Our approach considers the dependence of the cross section with the ion 
charge $Z$ and incorporates the scaling of the ionization with the 
number of active electron $n_e$ of the molecular targets. Scaling rules 
are generally very useful since they can be used as first-order 
approximations in experimental measurements and multipurpose codes. 

\section{Scaling rules}

\subsection{Scale with the ion charge}
\label{sec:zscaling}

In the development of our scaling rule, we examine forty collisional 
systems. The target-ion systems are composed of eight targets: 
the DNA and RNA nucleobases --adenine, cytosine, guanine, thymine, 
uracil--, tetrahydrofuran (THF), pyrimidine, and water; and five ion 
species: H$^+$, He$^{+2}$, Be$^{+4}$, C$^{+6}$, and O$^{+8}$. 
We consider these systems as a benchmark for the present rule. 

\begin{figure*}[!htb]
\centering
\includegraphics[width=0.9\textwidth]{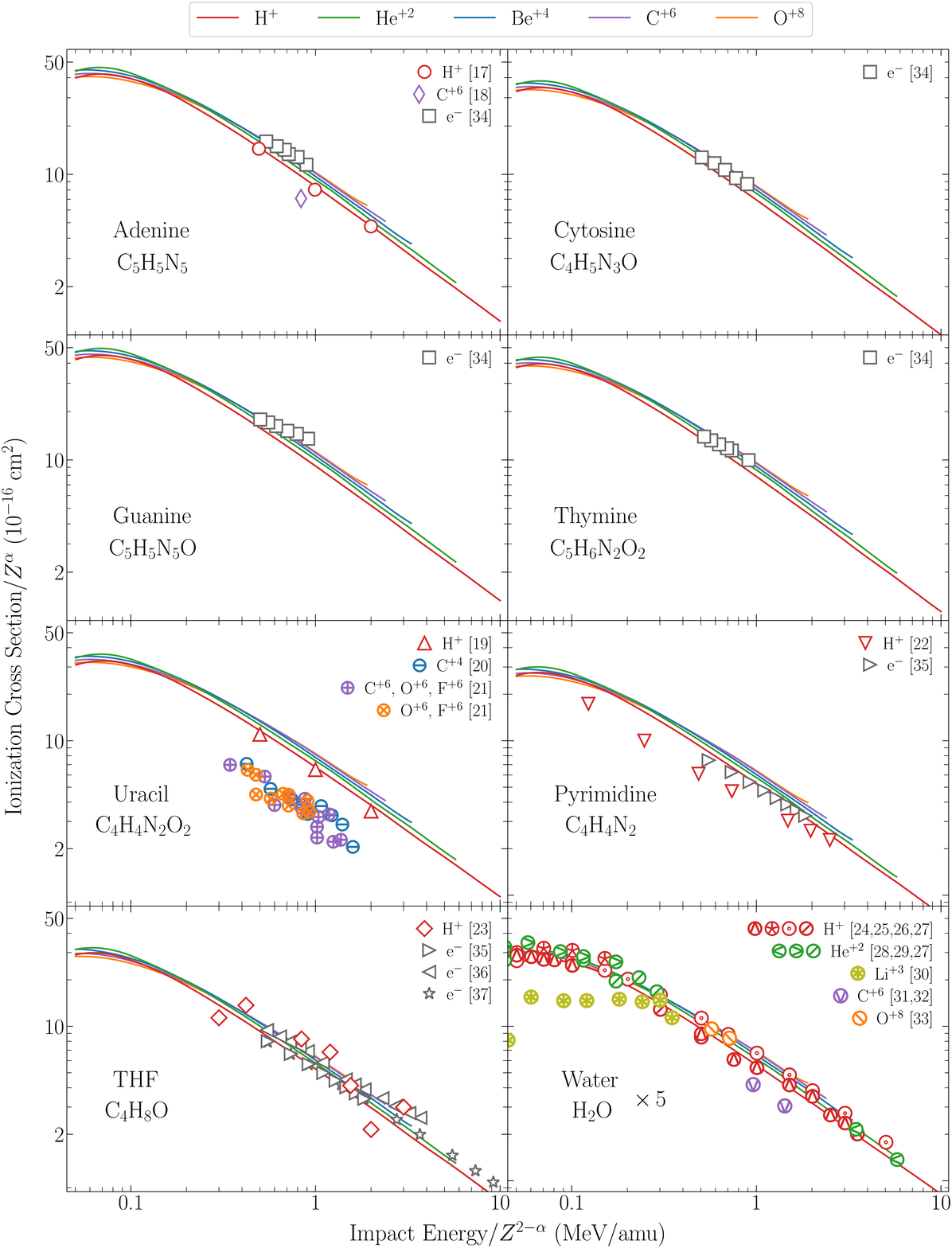}
\caption{(Color online) Scaled ionization cross section $\sigma/Z^{\alpha}$ 
as a function of the ion impact energy $E/Z^{2-\alpha}$ with $\alpha=1.2$. 
Colors are associated with the incident ion labeled on top of the figure. 
Curves: present CDW-SSM theoretical results. 
Symbols: experimental data~\cite{itoh2013,iriki2011,wolff2014,wang2016,
tribedi2019,agnihotri2012,agnihotri2013,Luna2007,Rudd86,pRudd85,
toburen80,Ohsawa05,Bhattacharjee17,Luna_Li_water,DalCappello2009,
Tribedi_O_water}. Electron impact ionization values~\cite{rahman2016,
bug2017,wolf2019,fuss2009} are included with the corresponding 
equi-velocity conversion.}
\label{fig:zreduced}
\end{figure*} 

We found two types of $Z$-scaling laws in the literature applicable 
to the intermediate impact energy range. The rule suggested by Janev 
and Presnyakov~\cite{janev1980} considers $\sigma/Z$ versus $E/Z$ to be 
the \textit{natural} reduced form of the ionization cross section 
$\sigma$ and the incident ion energy $E$. More recently, Montenegro and 
co-workers~\cite{dubois13,montenegro_pra13} suggested an alternative 
scaling by taking into account that the cross section is a function of 
$Z^2/E$ at high energies. Their scaling, given by 
\begin{equation}
 \sigma/Z^{\alpha}=f(E/Z^{2-\alpha}),
\label{eq:Montenegro}
\end{equation}
keeps the $Z^2/E$ relationship for any value of the parameter 
$\alpha$. The authors proposed $\alpha=4/3$ for the ionization of He and 
H$_2$ by differently charged ions~\cite{dubois13}. 
 
Following the work of Montenegro and collaborators, we found that the 
parameter that best converges the CDW-SSM cross sections of the forty 
collisional systems over the broadest energy range is $\alpha=1.2$. 
The validity of this particular scaling is evident in 
Fig.~\ref{fig:zreduced}, where --for each target-- the CDW-SSM curves
corresponding to different ions lay one over the other. 
It is worth noting that our theoretical results are valid for impact
energies above the maximum of the cross sections, which corresponds to 
an impact energy range from 50 keV for H$^+$ to 250 keV/amu for 
O$^{+8}$.

We also examined the experimental data available for the forty 
ion-target systems~\cite{itoh2013,iriki2011,wolff2014,wang2016,
tribedi2019,agnihotri2012,agnihotri2013,Luna2007,Rudd86,pRudd85,
toburen80,Ohsawa05,Bhattacharjee17,Luna_Li_water,DalCappello2009,
Tribedi_O_water} with the $Z^\alpha$-scaling rule. For targets with 
none or little experimental data, we included electron impact 
ionization results~\cite{rahman2016,bug2017,wolf2019,fuss2009} at high 
velocity with the corresponding equivelocity conversion. As can be 
noted, most of the data in Fig.~\ref{fig:zreduced} confirm the present 
scaling, even for O$^{+8}$ in water~\cite{Tribedi_O_water}. Only two 
data sets are off our predictions: the ionization cross section of 
uracil by swift C, O, and F ions from Refs.~\cite{agnihotri2012,
agnihotri2013}, and the values for Li$^{+3}$ in water from 
Ref.~\cite{Luna_Li_water} for $E<600$~keV/amu. In the case of uracil, 
recent CTMC calculations by Sarkadi~\cite{sarkadi2016} are also above 
the experimental values by Tribedi and collaborators~\cite{agnihotri2012,
agnihotri2013}.

\subsection{Scale with the molecular target}

The good results obtained in the scaling with the ion charge 
encouraged us to further investigate a scaling law that could 
predict values for ionization cross sections of any ion in any 
molecule. To this end, we considered the number of active electrons 
in each molecule $n_e$ proposed in Ref.~\cite{MendezJPB20} and combined 
it with the $Z^\alpha$-scaling from Section~\ref{sec:zscaling}.

In our previous work, we noticed that the CDW ionization cross sections 
$\sigma_A$ of atomic targets H, C, N, and O scale with the number of 
active electrons per atom $\nu_A$, as \mbox{$\sigma_e=\sigma_A/\nu_A$,} 
where $\nu_A$ is $1$ for H and $4$ for C, N, O, i.e.,
\begin{equation}
 \frac{\sigma_{\mathrm{H}}}{1}\sim
 \frac{\sigma_{\mathrm{C}}}{4}\sim
 \frac{\sigma_{\mathrm{N}}}{4}\sim
 \frac{\sigma_{\mathrm{O}}}{4}\,.
\end{equation}
By means of the SSM, we define the number of active electrons per 
molecule as $n_e=\sum_A n_A \nu_A$. The $n_e$ values for the 
molecular targets considered throughout this work are displayed in 
Table~\ref{nn}. The scaling with the molecular number of active
electrons proved to give excellent results, as shown in Fig.~6 of 
Ref.~\cite{MendezJPB20}.

\subsection{Scale with the ion charge and \\ the molecular target}

\begin{figure*}[!htb]
\centering
\includegraphics[width=0.8\textwidth]{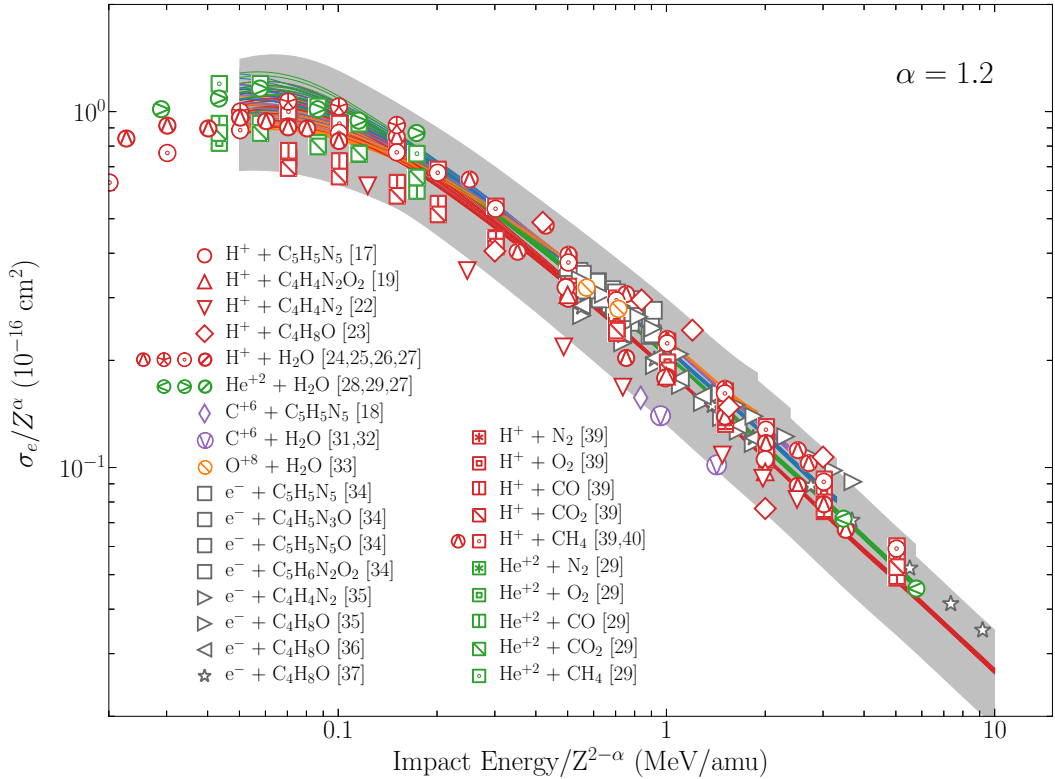}
\caption{(Color online) Ionization cross section reduced with the ion
charge $Z$ and scaled with number of active electrons per molecule $n_e$,
given by Eq.~(\ref{eq:u-scaling}) with $\alpha=1.2$. 
Curves: present CDW-SSM theoretical results. 
Symbols: experimental data~\cite{itoh2013,iriki2011,wolff2014,wang2016,
tribedi2019,Luna2007,Rudd86,pRudd85,toburen80,Ohsawa05,Bhattacharjee17,
DalCappello2009,Tribedi_O_water,Rudd85,Luna2019,Rudd1983}. Electron 
impact ionization values~\cite{rahman2016,bug2017,wolf2019,fuss2009} 
are included with the corresponding equi-velocity conversion.}
\label{fig:zalpha}
\end{figure*}

By incorporating the $Z^\alpha$ reduction and the scaling with the 
number of active electrons, we introduce the scaled and reduced 
ionization cross section of molecules $\tilde{\sigma}$, which is 
expressed as a function of $E/Z^{2-\alpha}$, and it is given by
\begin{equation}
 \tilde{\sigma}=\frac{\sigma_e}{Z^{\alpha}}=\frac{\sigma_M/n_e}{Z^{\alpha}}\,,
\label{eq:u-scaling}
\end{equation}
where $\sigma_M$ is the ionization cross section for the molecular 
target, $n_e$ is the number of active electrons per molecule displayed 
in Table~\ref{nn}, and the parameter is $\alpha=1.2$. 
Fig.~\ref{fig:zalpha} shows the theoretical and experimental values of 
$\tilde{\sigma}$ (given by Eq.~\ref{eq:u-scaling}) for all the systems 
displayed in Fig.~\ref{fig:zreduced}. As can be noted, the scaling 
works very well and is independent of the ion charge or the complexity 
of the molecular target. Our theoretical curves lay in a narrow band 
valid for any charged ion (reduced with $Z^\alpha$) in any molecule 
(scaled with the number of active electrons) with a dispersion of about 
$\pm 20\%$. If we consider the experimental data, the uncertainty of 
our scaling grows to $\pm 30\%$, which is schematized in 
Fig.~\ref{fig:zreduced} with a gray area. It is worth noting that we 
did not include in this figure the data for uracil from 
Refs.~\cite{agnihotri2012,agnihotri2013}, and Li$^{+3}$ on 
water~\cite{Luna_Li_water}. The discussion about these experimental 
values exceeds the present work.

We consider the present scaling robust enough to be valid for different 
ion-molecule combinations. We tested the generality of our model 
by including in Fig.~\ref{fig:zreduced} several data sets of 
molecular targets not considered previously, such as the measurements 
by Rudd~\textit{et al.}~\cite{Rudd85,Rudd1983} for H$^{+}$ and He$^{+2}$ 
in N$_2$, O$_2$, CH$_4$, CO and CO$_2$, and recent values by 
Luna~\textit{et al.} \cite{Luna2019} for H$^{+}$ in CH$_4$ . 

The good agreement shown in Fig.~\ref{fig:zalpha} summarizes the main 
result of this work and holds the validity of the present scaling for 
different ions and targets. Although the theoretical CDW-SSM results 
are valid for energies above the maximum of the cross sections, the 
scaling of the experimental data extends even to lower impact energies, 
as can be noted in Fig.~\ref{fig:zalpha}. New measurements for other 
ions and molecules are expected to reinforce the present proposal. 

\begin{table}[t]
\begin{center}
\begin{tabular}{|ll|ll|ll|}
\hline
 Molecule & $n_e$ & Molecule          & $n_e$ & Molecule          & $n_e$ \\
\hline
 H$_2$O   & 6  & CO$_2$               & 12 & C$_4$H$_5$N$_3$O     & 37   \\ 
 N$_2$    & 8  & C$_4$H$_8$O          & 28 & C$_5$H$_6$N$_2$O$_2$ & 42   \\ 
 O$_2$    & 8  & C$_4$H$_4$N$_2$      & 28 & C$_5$H$_5$N$_5$      & 45   \\ 
 CH$_4$   & 8  & C$_4$H$_4$N$_2$O$_2$ & 36 & C$_5$H$_5$N$_5$O     & 49   \\ 
 \hline
\end{tabular}
\caption{Number of active electrons per target at intermediate to high 
energies obtained from the CDW calculations~\cite{MendezJPB20}.}
\label{nn}
\end{center}
\vspace{-0.75cm}
\end{table}

\section{Conclusions}

We present scaling rules for the ionization
cross sections of highly charged ions in biological targets. The first
scaling reduces the nature of the projectile by scaling the cross 
section with the ion charge, $Z^{\alpha}$, as a function of the reduced 
impact energy $E/Z^{2-\alpha}$, with $\alpha=1.2$. The second scaling 
considers the molecular description of the target by taking into 
account the number of active electrons per molecule, $n_e$. The last 
scaling law combines the $Z^{\alpha}$-reduction with the $n_e$-scaling 
of the cross section, and it becomes independent of the ion charge and the 
molecular target. The scalings are obtained by means of CDW-SSM 
calculations for five different charged ions in eight targets and 
tested with the available experimental data. The generality of our 
independent scaling is proved to be valid in a wide energy range by 
considering a significant number of experimental data sets for other 
collisional systems.

\section{Acknowledgments}

This work was finantially supported by Consejo Nacional de
Investigaciones Cient\'ificas y T\'ecnicas (PIP 2014), Agencia 
Nacional de Promoci\'on Cient\'ifica y Tecnol\'ogica (PICT 2017-2945), 
and Universidad de Buenos Aires (UBACyT 20020170100727).

\end{document}